\documentclass[9pt, a4paper]{article}

\usepackage{anysize}
\marginsize{50pt}{40pt}{0pt}{60pt}
\usepackage{cuted}  
\usepackage{placeins}
\usepackage[version=4]{mhchem}
\usepackage{siunitx}
\DeclareSIUnit\Molar{M}
\usepackage[utf8]{inputenc}
\usepackage{graphicx}
\usepackage{xcolor}
\usepackage{array}
\usepackage{multirow}
\usepackage{float}
\usepackage{amsmath}
\usepackage{amsfonts}
\usepackage{amssymb} 
\usepackage{bbm}
\usepackage{microtype}
\usepackage{booktabs}
\usepackage{abstract} 
\usepackage{lipsum}
\usepackage{siunitx}  
\makeatletter
\@ifundefined{@setmarks}{\let\@setmarks\relax}{}
\makeatother

\usepackage[obeyFinal]{todonotes}

    \newcommand{\todoAfterSubmission}[1]{}

{\begin{strip}%
    \begin{center}%
    \begin{minipage}{0.9\textwidth}
      \begin{center}%
        {\bfseries Significance Statement}%
      \end{center}%
      \medskip%
}%
{\end{minipage}%
    \end{center}%
\end{strip}%
\medskip}%

\renewenvironment{abstract}%
{\begin{strip}%
 \begin{center}%
    \begin{minipage}{0.9\textwidth}
      \begin{center}%
        {\bfseries Abstract}%
      \end{center}%
      \medskip%
}%
{\end{minipage}%
    \end{center}%
\end{strip}%
}%

\newcommand{\keywordsfont}{\normalfont\rmfamily\fontsize{7}{10}\selectfont}

\usepackage{caption}
\renewcommand{\floatpagefraction}{0.95} 

\usepackage[printonlyused]{acronym}
\acrodef{SNN}{Spiking Neural Network}
\acrodef{TTFS}{time-to-first-spike}
\acrodef{ANN}{Artificial Neural Network}

\definecolor{CC_blue}{RGB}{16, 38, 148}  

\usepackage[                
        backend=bibtex,      
        style=authoryear,      
        natbib=true,        
        hyperref=true,      
        alldates=year,      
	    uniquelist=false,   
        sorting=nyt,
	    isbn=false,
	    eprint=false,
	    doi=false,
        url=false,
	    maxcitenames=2,
        giveninits=true,
        uniquename=init
]{biblatex}

\renewbibmacro{in:}{%
  \ifentrytype{article}{}{%
    \printtext{\bibstring{in}\intitlepunct}%
  }
}

\AtEveryBibitem{
    \clearfield{urlyear}
    \clearfield{urlmonth}
    \clearlist{language}
}

\newcommand{\methref}[1]{Methods: \nameref{#1}}
\newcommand{\SIref}[1]{Supplemental Information: \autoref{#1}}

\usepackage[colorlinks=true,
            urlcolor=CC_blue,
            citecolor=CC_blue,
            linkcolor=CC_blue,
            bookmarks=true,
            linktoc=all,
            breaklinks]{hyperref}

\usepackage{orcidlink} 

\newcommand{\tikzHex}[1][black,fill=white]{\tikz[baseline=-0.1cm] 
   \newdimen\R
   \R=0.12cm
   \draw[#1] (0:\R) \foreach \x in {60,120,...,360} {  -- (\x:\R) };%
}%
\newcommand{\tikzHexTurned}{\tikzHex[rotate=30]}

\begin{document}
\twocolumn[
    {\centering{{\Huge Synchronization and semantization\\in deep spiking networks \par}\vspace{3ex}
    
    { \large
      Jonas Oberste-Frielinghaus\textsuperscript{1,2*}\orcidlink{0009-0000-5582-0549}, Anno C. Kurth\textsuperscript{1,3}\orcidlink{0000-0002-9557-1003}, Julian Göltz\textsuperscript{4,5}\orcidlink{0000-0001-9170-4422}, Laura Kriener\textsuperscript{6,5}\orcidlink{0000-0001-5275-9199}, \\Junji Ito\textsuperscript{1}\orcidlink{0000-0001-9170-4422},  Mihai A. Petrovici\textsuperscript{5}\orcidlink{0000-0003-2632-0427}, Sonja Grün\textsuperscript{1,7,8}\orcidlink{0000-0003-2829-2220}
      \par
    }
      
    {
      \newcounter{affilcounter}
      \setcounter{affilcounter}{1}
      \begin{list}{$^{\text{\arabic{affilcounter}}}$}{\usecounter{affilcounter}
      \setlength{\labelsep}{0.1cm}
      \setlength{\labelwidth}{0.5cm}}
        \item Institute for Advanced Simulation (IAS-6), Jülich Research Centre, Jülich, Germany\vspace{-6pt}
        \item RWTH Aachen University, Aachen, Germany\vspace{-6pt}
        \item RIKEN Center for Brain Science, Wako, Saitama, Japan\vspace{-6pt}
        \item Kirchhoff-Institute for Physics, Heidelberg University, Heidelberg, Germany\vspace{-6pt}
        \item Department of Physiology, University of Bern, Bern, Switzerland\vspace{-6pt}
        \item Institute of Neuroinformatics, University of Zurich and ETH Zurich, Zurich, Switzerland\vspace{-6pt}
        \item JARA Brain Institute I (INM-10), Jülich Research Centre, Jülich, Germany\vspace{-6pt} 
        \item Theoretical Systems Neurobiology, RWTH Aachen University, Germany\vspace{-6pt}
      \end{list}}
      
      \raggedright \quad * j.oberste-frielinghaus@fz-juelich.de \par
      \medskip
      \keywordsfont keywords: spiking neural networks, network dynamics, computation, latency coding, synchrony, artificial intelligence, training algorithm \vspace{-15pt}
    }
    }
]

\begin{abstract} 
    Recent studies have shown how spiking networks can learn complex functionality through error-correcting plasticity, but the resulting structures and dynamics remain poorly studied.
    To elucidate how these models may link to observed dynamics in vivo and thus how they may ultimately explain cortical computation, we need a better understanding of their emerging patterns.
    We train a multi-layer spiking network, as a conceptual analog of the bottom-up visual hierarchy, for visual input classification using spike-time encoding.
    After learning, we observe the development of distinct spatio-temporal activity patterns.
    While input patterns are synchronous by construction, activity in early layers first spreads out over time, followed by re-convergence into sharp pulses as classes are gradually extracted.
    The emergence of synchronicity is accompanied by the formation of increasingly distinct pathways, reflecting the gradual semantization of input activity.
    We thus observe hierarchical networks learning spike latency codes to naturally acquire activity patterns characterized by synchronicity and separability, with pronounced excitatory pathways ascending through the layers.
    This provides a rigorous computational hypothesis for the experimentally observed synchronicity in the visual system as a natural consequence of deep learning in cortex.
    
    \medskip
    \begin{center}%
        {\bfseries Significance Statement}%
    \end{center}%
    \medskip
    Recent advances in AI have rekindled the hypothesis of deep learning in the brain, but there remains a significant gap at the microscopic scale, as cortical neurons communicate with sparse and discrete signals, rather than continuously in time.
    Building on an analytical model of deep learning with spikes, we investigate the emergence of spatio-temporal structures in hierarchical spiking networks.
    We find that neuronal populations learn to form tight pulse packets for downstream communication and observe distinct pathways of neuronal excitation that become increasingly separated with network depth, indicating the progressive semantization of neuronal activity.
    This puts forth a rigorous computational hypothesis for the well-established experimental observations of synchrony and semantization in sensory cortex.
\end{abstract} 

\section*{Introduction}

Artificial neuronal networks (ANNs) are the backbone of modern machine learning applications.
Since the formulation of the perceptron \parencite{Rosenblatt58_386}, ANNs have gradually diverged away from the biology that originally inspired them, but their recent success across many domains has prompted a broad interest to reevaluate their applicability as models of processing in the brain \parencite{Richards19_1761}.
Many of these studies focus on visual processing, as it is among the best studied computational tasks, both in cortex and as an application for AI.
As an example, Convolutional Neural Networks \parencite{Lecun98_2278, Krizhevsky12_1097} are used successfully as model of the visual system \parencite{Yamins16_356, Lindsay21_2017}.

However, the underlying models remain very close or even identical to conventional ANNs, in particular by using continuous neuronal transfer functions.
This is markedly different from cortical networks, in which inter-neuron communication is dominated by action potentials, or spikes, i.e., cortical networks are spiking neural networks (SNNs).
A continuous transfer function can be approximated in SNNs by considering the average spike rates over time or populations of neurons, leading to an interpretation that the aforementioned models are operating in a purely rate coding framework.
Even though rate coding has been highly influential in Neuroscience, may it be for characterizing response properties of single neuron \parencite{Hubel62, Georgopoulos82_1527} or neural populations (from population rates \parencite{Georgopoulos86_1416, Churchland12_51} to geometric interpretation of the evolution of the population vector \parencite{Gao17_4262, Gallego17_978, Stringer19_361, MoralesGregorio24_114371}), rate coding is far from the only operational mode of the cortex.
Alternative, well-established computational interpretations of cortical activity emphasize the fine temporal nature of neural activity, e.g., \parencite{Abeles91, Thorpe01_715, Izhikevich06_245}.
They are supported by experimental findings such as the coordinated spiking on millisecond scale \parencite{Riehle97_1950, Prut98, Kilavik09_12653, Torre16_8329} or characteristic temporal sequences of spikes \parencite{Yiling23_3021, Xie24_935, Sotomayorgomez25_115547}.

The main reason for using rate-based models, i.e., models that only communicate via firing rates emulating a continuous transfer function and not precise spikes, lies in the difficulty of training SNNs.
Indeed, it is not obvious how to calculate gradients of discrete spiking activity, which would be necessary for a straightforward application of error backpropagation.
However, recent years have seen the development of various approaches capable of overcoming this challenge, most notably approximate surrogate methods \parencite{Neftci17_324, Zenke18_1514, Yin23_518} and exact spike-time gradients \parencite{Bohte02_17, Wunderlich21_12829, Goeltz21_823}.
These now allow the training of deep spiking networks to performances comparable with their conventional counterparts.
Thus, such networks can form the basis of a more rigorous reassessment of the deep learning hypothesis in the brain, now also taking into account a more realistic form of spike-based, as opposed to continuous, communication.

With trained networks it is possible to study how their structure and activity is shaped through learning and which characteristic patterns emerge.
In particular, the aspects of propagation and transformation of the neural code \parencite{Perkel68} and their underlying mechanisms can be investigated thoroughly.
There have been extensive studies about the propagation of activity in SNNs, e.g., in simulations \parencite{Diesmann99_529, VanRossum02, Vogels05_10786} or in vitro \parencite{Reyes03, Barral19_3969}, but the studied networks were not trained to perform a particular task.

Here, we consider multi-layered SNNs trained by exact gradient descent as visual image classifiers using a spike latency code \parencite{Goeltz21_823}.
Thereby we approach their activity as we would approach electrophysiological recordings, but with the added benefit of having access to all observables in the network, as opposed to the massive subsampling that is characteristic of in-vivo data \parencite{Levina22_770}.
In the following, we show how these networks form very distinct activity and connectivity patterns.
In particular, we show that neuron subpopulations in these networks learn to synchronize their firing in response to patterns of a particular class.
This is a phenomenon frequently observed in the cortex, e.g., \parencite{Gray89_1698, Gray89_334}), but here we show that it arises from learning by gradient descent, thus providing a functional explanation.
Moreover, we observe how these populations grow increasingly distinct across the network hierarchy, demonstrating the semantization of activity as it propagates downstream.
This bundling of activity in space and time maps closely to various experimental observations, thus establishing a first step towards a rigorous link between the theory of learning by gradient descent in spiking networks and in-vivo recordings of cortical activity.

\section*{Results}

\subsection*{Activity in the network}\label{sec:propagation}

{\renewcommand{\floatpagefraction}{0.9}
\begin{figure*}[t]
    \centering
    \includegraphics[width=\textwidth]{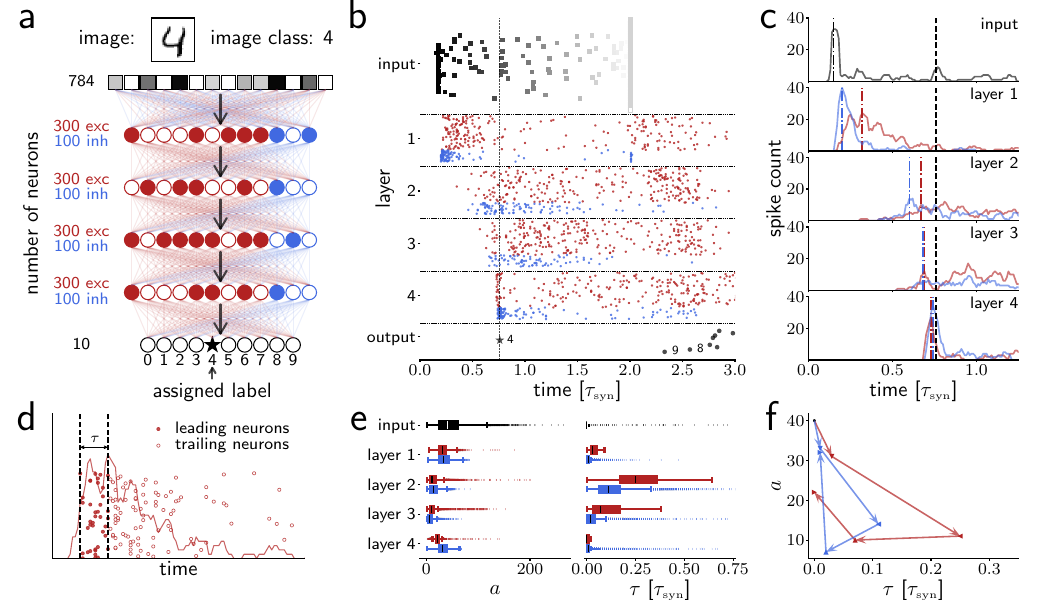}
    \caption{\textbf{Propagation of activity through the network.}
        (\textbf{a}) Structure of the network and the task.
        One randomly chosen image of class 4 is passed into the network via the input layer.
        Each pixel is represented by a neuron ($\num{784}$ in total).
        The input propagates through four hidden layers with $\num{300}$ excitatory and $\num{100}$ inhibitory neurons.
        The output layer contains 10 output neurons, each representing one class.
        The network assigns a label according to the neuron in the output layer that spikes first.
        (\textbf{b}) Activity in response to the image in \textbf{a} as raster plot.
        Time is shown as multiples of the synaptic time constant $\tau_{\text{syn}}$, so all our conclusions remain scale-invariant with respect to the specific time constants in the network.
        The dots represent the spike times of the individual neurons.
        The image is represented by a latency code via spike times in the input layer.
        The brightness of the dot corresponds to the brightness of the pixel in the image (the darker, the earlier).
        In the hidden layers (1-4) red dots correspond to excitatory, blue dots to inhibitory neurons.
        The first spike in the output layer is marked by an asterisk.
        The image is classified correctly as ``4''.
        (\textbf{c}) Spike time histogram for the activity in \textbf{b}.
        The spike count was measured in a sliding window of $0.05\:\tau_{\text{syn}}$ in $0.01\:\tau_{\text{syn}}$ steps.
        In hidden layer 1-4 the spike count is separated between excitatory and inhibitory neurons.
        The colored vertical lines as well as the first black line in the panel for the input layer denote the maximum of the histogram before the first output spike (dashed black line).
        (\textbf{d}) Illustration of the characterization of the activity.
        We determine the rise time $\tau$ as the time from the first spike to the maximum of the histogram, we term neurons active during this time \textit{leading} neurons and the others \textit{trailing} neurons.
        $a$ is the number of leading neurons.
        (\textbf{e}) Box plots of the distributions of $a$ (left) and $\tau$ (right) across all images, separate for excitatory and inhibitory neurons in the hidden layers. 
        The line marks the median, the box marks the range between the first (Q1) and the third quantile (Q3), the whiskers range from the box to the lowest data point above $\text{Q1} - 1.5(\text{Q3}-\text{Q1})$ and the highest data point below $\text{Q3} + 1.5(\text{Q3}-\text{Q1})$, fliers represent outliers.
        (\textbf{f}) State space representation of the medians of the distributions in \textbf{e} in sequence of the layers.
        Arrows point in the direction of the propagation of activity in the network.
    }
    \label{fig:network_activity}
\end{figure*}
}

We investigate a feed-forward network with all-to-all connections between consecutive layers consisting of an input, four hidden, and an output layer as depicted in \autoref{fig:network_activity}a.
As a classical visual benchmark that does not require complicated structures lacking direct biological equivalents, such as perfect copies of convolutional kernels or max-pooling layers, we chose classification of the MNIST dataset \parencite{Lecun98_2278} as task for the network.
Importantly, and unlike in classical ANNs, the neurons in the hidden layers obey Dale's law \parencite{Eccles57}, meaning that each neuron has either only excitatory or only inhibitory outgoing connections.
To roughly approximate the ratio found in cortex \parencite{Markram04}, each hidden layer consists of $\num{300}$ excitatory and $\num{100}$ inhibitory neurons.
The output layer has 10 neurons, one for each image class.

To understand how the network processes the inputs, we first examine how spiking activity propagates through the layers in response to an arbitrary image of a handwritten digit (see \autoref{fig:network_activity}a) after training (\autoref{fig:network_activity}b).
The input image is converted into a set of spike times that specify the activity of the input layer.
Each neuron in the input layer corresponds to a pixel of the input image, the brightness of which determines whether the corresponding neuron fires earlier or later; the darker the pixel, the earlier the neuron fires.
The spiking activity of the input layer is passed to the subsequent layer (layer 1), where incoming spikes influence the membrane potential of the neurons.
If, for a given neuron in layer 1, the evoked membrane potential exceeds the threshold, the neuron emits a spike that is passed to all neurons in the next layer (layer 2), and so on.
Typically, to bring a neuron to fire, it needs to receive sufficient synchronous input from the preceding layer.
This way, spiking activity propagates downstream, from layer to layer.
In the output layer, the label assigned to the input is determined by which of the 10 output neurons emits a spike first.
On the test dataset, the trained network achieves an accuracy of 0.98, i.e, the assigned label matches the image class of the input for 98\% of the test images.

To examine the propagation of the activity quantitatively, we first calculate the spike time histogram for the spiking activity shown in \autoref{fig:network_activity}b.
Starting in the input layer, we observe a sharp peak in the histogram (\autoref{fig:network_activity}c first row).
Over the next two layers, the spike times spread out, and hence the histogram peak becomes less prominent.
Then in layer 3, the excitatory neurons synchronize again, ultimately resulting in a very sharp peak for both the excitatory and inhibitory neurons in layer 4.
Overall, these activity profiles resemble the propagation of a \textit{pulse packet} i.e., a synchronous volley of spikes, through the layers, which first disperses and then re-synchronizes over the layers.

We characterize the pulse packets of excitatory and inhibitory neurons in each layer individually by, on the one hand, determining the rise time $\tau$ of the spike time histogram, i.e., the time from the first spike to the maximum of the histogram.
The rise time gives an estimate of how synchronous the spikes occur in the layer.
On the other hand, we count the number $a$ of neurons that fire spikes during this time (see \autoref{fig:network_activity}d).
These neurons we term ``leading neurons'', and the neurons that fire after this period we term ``trailing neurons''.

\autoref{fig:network_activity}e shows in the form of box plots the distributions of the number of leading neurons ($a$, left) and the rise time of the histogram ($\tau$, right) across all images.
Over all images the same trend as we observe in the example shown in \autoref{fig:network_activity}c solidifies; the activity starts with a high and sharp peak (large $a$ and small $\tau$), then gets dispersed (smaller $a$ and larger $\tau$), and then builds up again (see \autoref{fig:network_activity}f).
This trend is evident for both the excitatory and the inhibitory neurons.

In summary, we see that the propagation of activity in the network is characterized by a pulse packet that decays and then builds up again.
The conditions for networks to exhibit this kind of activity have been investigated extensively \parencite{Diesmann99_529, Tetzlaff02_673, Vogels05_10786, Kumar08_1, Shinozaki10_011913}.
More on this point will follow in the discussion.

While the characterization of the activity as a pulse packet allows a quantitative description of the activity propagation through the layers, it does not immediately provide functional implications of the observed activity for information processing.
Assuming that the pulse packet plays a relevant role in achieving a correct classification, for a pulse packet to represent the image class of the input, it would need to encode the information by the identity of the neurons that contribute spikes to it.
Since we observe that the pulse packet is gradually built up as it propagates through the layers, we also expect that its representation of the image class would be progressively consolidated towards deeper layers, i.e., a more specific subset of neurons would provide spikes to the pulse packet in deeper layers.
This leads us to a close examination of the identity of the leading neurons, as shown in the following section.

\subsection*{Representation of classes in the activity}\label{sec:representation}

\begin{figure*}[t]
    \centering
    \includegraphics[width=\textwidth]{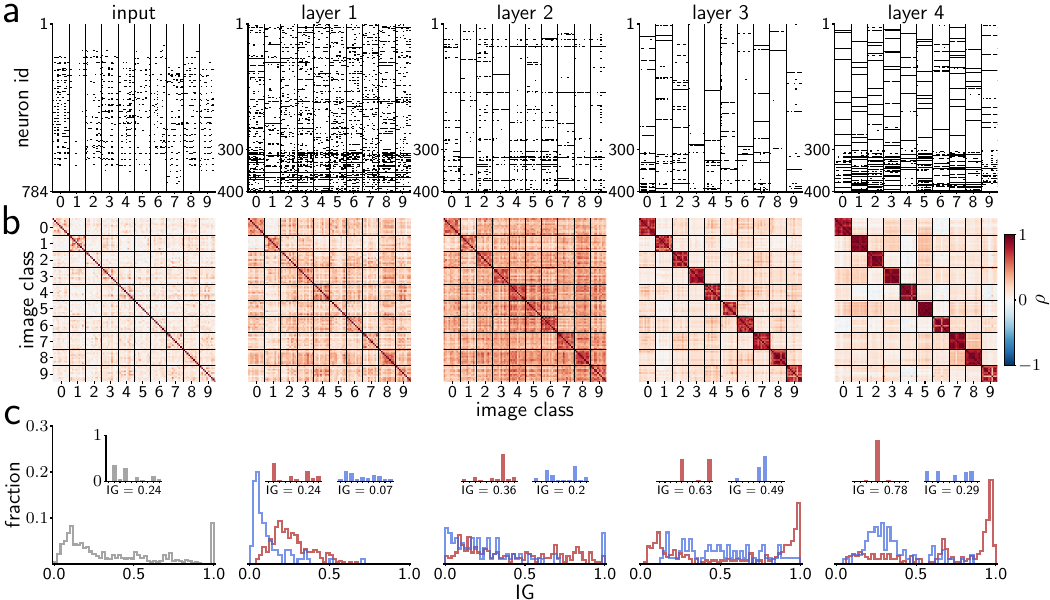}
    \caption{\textbf{Representation of labels across the layers.}
             (\textbf{a}) State of activity for all neurons for 100 images (10 randomly selected for each of the 10 classes), ordered according to the classes.
             If the neurons is a leading neuron for the image it is marked black, if it is a trailing neuron it is marked white (for the definition see \autoref{fig:network_activity}d).
             For the hidden layers (1-4) the inhibitory neurons are indexed with numbers between 301 and 400, the first 300 being excitatory neurons.
             (\textbf{b}) Matrices of similarities $\rho$ of any two leading neuron sets shown in \textbf{a}, ordered by image class, and color-coded (see colorbar on the right).
             (\textbf{c}), Distributions of the specificity of all neurons per layer measured by the information gain (IG) and normalized by the number of neurons (neurons that are never active are excluded).
             On the very left this distribution is shown for the input neurons, then for the hidden layers (1-4) (left to right), separately for the excitatory (red) and inhibitory (blue) neurons.
             To illustrate the meaning of the IG, for each population, we choose a neuron with the median IG and plot the distribution of the image classes of the images if the neuron was a leading neuron in the inset.
    }
    \label{fig:representation}
\end{figure*}

Next, we investigate how different classes are represented in the population activity of each layer.
We focus on the leading neurons here because these neurons are likely most important for the classification, since the network operates on a latency code and these neurons fired spikes with the shortest latencies in the individual layer.
Thereby we consider a set $\mathcal{V}^l_{x}$ of the leading neurons in layer $l$ for image $x$ (see \methref{sec:meth_active} for details).

\autoref{fig:representation}a shows, separately for each layer, the leading neurons $\mathcal{V}^l_{x}$ for 100 randomly chosen test images (10 for each class).
While in the early layers no particular structure can be discerned, in the deeper layers certain neurons fire across all images of a particular class, forming a bar-code-like pattern.
We note that these observations are not dependent on our specific way of defining $\mathcal{V}^l_{x}$; other equally plausible definitions of $\mathcal{V}^l_{x}$ lead to essentially identical observations (see \SIref{fig:SI_other_def}).

To quantify the consistency of the leading neurons in response to different images of the same class, we calculate the similarity $\rho^l_{x, y}$ of the leading neuron sets $\mathcal{V}^l_{x}$ and $\mathcal{V}^l_{y}$ in layer $l$ for two respective images $x$ and $y$ as (\autoref{fig:representation}b):
\begin{equation}
    \rho^l_{x, y} = \frac{N^l|\mathcal{V}^l_{x} \cap \mathcal{V}^l_{y}| - |\mathcal{V}^l_{x}||\mathcal{V}^l_{y}|}{\sqrt{|\mathcal{V}^l_{x}||\mathcal{V}^l_{y}|(N^l - |\mathcal{V}^l_{x}|)(N^l - |\mathcal{V}^l_{y}|) }}\;,
    \label{eq:correlation}
\end{equation}
where $N^l$ is the number of neurons in layer $l$ and $|\mathcal{V}|$ denotes the cardinality of the set $\mathcal{V}$ (see \methref{sec:meth_corr} for details).
If the two sets are identical, $\rho^l_{x, y}=1$; if the activity is maximally dissimilar (which would be the case if half of the neurons were leading neurons for image $x$ and the other half for image $y$), $\rho^l_{x, y}=-1$; $\rho^l_{x, y}=0$ implies chance overlap.
\autoref{fig:representation}b shows the similarity calculated for all pairs of images used in \autoref{fig:representation}a for all layers, again grouped by the image classes.
Diagonal blocks correspond to similarities between images from the same class, while off-diagonal blocks quantify the similarity of the activity for images of different classes.
In the input layer and hidden layer 1, there is little difference between within-class and between-class similarities.
In layer 2, the degrees of the similarities within the diagonal blocks are higher than those in off-diagonal blocks, implying that images of the same class evoke more consistent activity than images of different classes.
This trend solidifies as activity propagates across layers, reaching its maximum in layer 4, where neural representations of images from the same class are almost identical.
The distribution of the overlap measures calculated for all pairs of the test images (\SIref{fig:SI_corr_dist}) confirms that this observation is not only for the 100 images randomly chosen here, but generally applies to all images.

To quantify the specificity of individual neurons in the representation of different image classes, we evaluate the Information Gain (IG) of a neuron, i.e., the information about the class of an input image gained by finding that neuron as a leading neuron for that image (for details see \methref{sec:meth_IG}).
An IG of $0$ implies that the neural firing is independent of the class of the input image; an IG of $1$ signifies that the neuron is fully indicative of a specific class.
\autoref{fig:representation}c shows the distributions of IGs across all neurons of the respective layers, separately shown for excitatory (red) and inhibitory (blue) neurons, excluding neurons that were not a leading neuron for any image.
In the input layer, we have a broad distribution of IGs with one peak roughly at $0.1$ and another at $1.0$. 
The latter represents the neurons that are leading neurons for exactly one image, thus being fully indicative of the class of that image.
The IG distributions for excitatory neurons in the hidden layers shift more and more towards an IG of $1.0$ in the deeper layers.
In contrast, IGs of inhibitory neurons are generally low and do not grow towards deeper layers, indicating that inhibitory neurons are less specific for one particular image class than excitatory neurons throughout the layers.
This is visualized by the two examples of the image class distribution for the neurons with the median IG in the respective layer and subpopulation (\autoref{fig:representation}c, inset) for all images when the neuron was a leading neuron.
For example, the excitatory neuron in the last layer is almost exclusively active for images of class 4, while the activity of the inhibitory neuron is more broadly distributed.
This is consistent with the findings in the visual cortex, where inhibitory neurons are more broadly tuned than excitatory neurons \parencite{Sohya07_2145, Niell08_7520, Lundqvist10_e803}.

\subsection*{Connectivity structure and path identification}\label{sec:path}

\begin{figure*}[t]
    \centering
    \includegraphics[width=\textwidth]{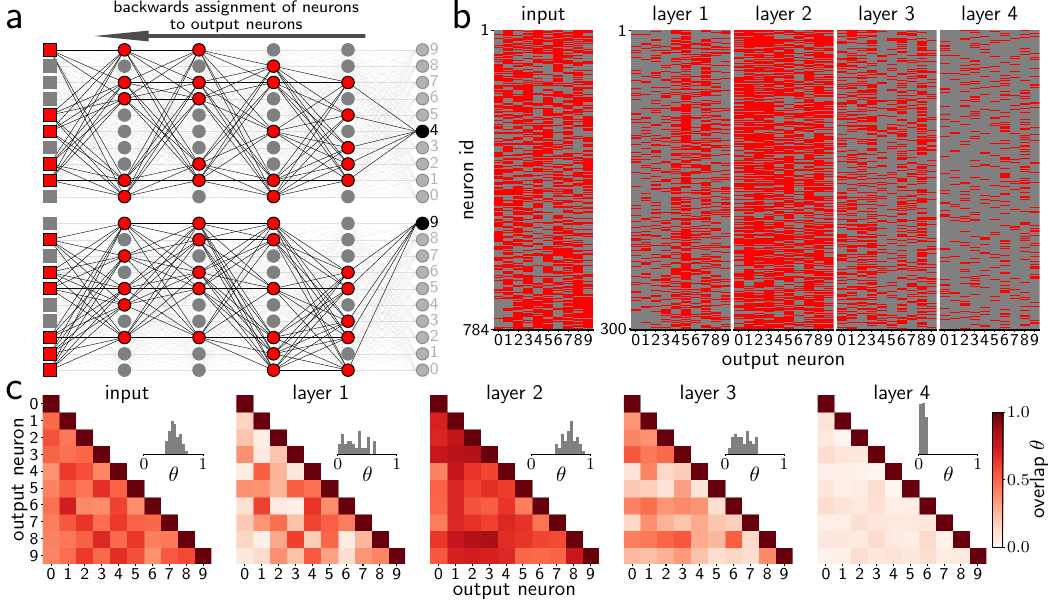}
    \caption{\textbf{Connectivity structure for the separation of image classes.}
             (\textbf{a}) Sketch for two identified paths, in the upper row for output neuron 4 in the lower for output neuron 9.
             In the upper row, neurons in the path i.e., in $\mathcal{P}_4$ are marked in red and neurons not in the path i.e., in $\mathcal{N}_4$, are marked grey.
             The neurons in $\mathcal{P}_4$ have stronger connections to the neurons within the path converging on output neuron 4.
             The path is identified by tracing the connections to output neuron 4 backwards through the network (see \methref{sec:meth_subset}).
             The same holds for the lower row for output neuron 9, mutatis mutandis.
             Each neuron can take part in multiple paths i.e., be part of  $\mathcal{P}_4$ and  $\mathcal{P}_9$ simultaneously.
             (\textbf{b}) Assignment of neurons to paths denoted by the identity of their respective label neurons on the abscissa.
             (\textbf{c}) Pairwise overlap between stages of the paths (\autoref{eq:similarity}).
             The degree of overlap is displayed in the colorbar.
             The insets show the distribution of overlap scores between pairs of pathways belonging to different output neurons (i.e., of the off diagonal elements of the presented matrices).
             }
    \label{fig:overlap}
\end{figure*}

So far we have concentrated on the neural activity, disregarding the knowledge of synaptic weights in the network.
We now turn to the synaptic weights and ask if we can find a relation between the connectivity structure of the network and the specificity of the neural activity.
In particular, we aim to identify neurons that have strong (direct or indirect) synaptic impacts on a specific output neuron.
To this end, we focus only on excitatory neurons, since the high specificity in the representation of image classes was found almost exclusively for excitatory neurons.

Our procedure for connectivity structure analysis is schematically illustrated in \autoref{fig:overlap}a.
We start by considering one specific output neuron $o$ ($o=4$ and $9$ in \autoref{fig:overlap}a top and bottom, respectively).
Then we identify neurons in the last hidden layer that are stronger connected to this output neuron $o$ than to the other output neurons.
The identified neurons (marked in red in \autoref{fig:overlap}a) constitute a subset $\mathcal{P}^4_{o}$ of excitatory neurons in layer $l=4$ with positive impact on the output neuron $o$, and complementarily, all the other excitatory neurons in layer 4 (marked in gray) are grouped into a subset $\mathcal{N}^4_{o}$ of neurons that do not have positive impact (for details see \methref{sec:meth_subset}).
In a similar manner, the subset $\mathcal{P}^3_{o}$ for layer 3 is defined by the excitatory neurons that preferentially target neurons in $\mathcal{P}^4_{o}$, and the subset $\mathcal{P}^3_{o}$ by all the other excitatory neurons in layer 3.
This procedure is repeated upstream through the whole network, and also for the other output neurons, defining $\mathcal{P}^l_{o}$ and $\mathcal{N}^l_{o}$ for all layers $l$ and all output neurons $o$.
Combing the subsets of neurons from all layers, we obtain a path $\mathcal{P}_{o} = \cup_{\forall l}\:\mathcal{P}^l_{o}$ through the network for each output neuron $o$, as well as a set of neurons not included in the path $\mathcal{N}_{o} = \cup_{\forall l}\:\mathcal{N}^l_{o}$.
Accordingly, we call the subsets $\mathcal{P}_o^l$ stages of the path $\mathcal{P}_{o}$.
Note that our construction of the paths is based solely on the connection preference of neurons for an output neuron, irrespective of the neural activity.
At the end, for each output neuron, a ``path'' through the network is identified, along which the neurons strongly influence the output neuron.

\autoref{fig:overlap}b shows the resulting subsets $\mathcal{P}^l_{o}$ and $\mathcal{N}^l_{o}$ for all 10 output neurons for all layers, where neurons in $\mathcal{P}^l_{o}$ are illustrated in red and neurons in $\mathcal{N}^l_{o}$ in gray.
In layer 4 we observe fewer neurons in $\mathcal{P}^4_{o}$ than $\mathcal{N}^4_{o}$ for all output neurons, in contrast to, e.g., layer 2 where much more neurons are in $\mathcal{P}^2_{o}$ than in $\mathcal{N}^2_{o}$.
At first, the paths become denser up until layer 2, where many neurons participate in different paths.
Then, from layer 2 to layer 4, the paths become increasingly sparse, such that fewer and fewer neurons contribute to the path to each individual output neuron.

To quantify how these sets of neurons become more specific to a particular output neuron in deeper layers, we calculate their pairwise overlap $\theta^l_{i, j}$ for all combinations of output neurons, akin to the cosine similarity of vectors: 
\begin{equation}
    \theta^l_{i,j} = \frac{|\mathcal{P}^{l}_{i} \cap \mathcal{P}^{l}_{j}|}{\sqrt{|\mathcal{P}^{l}_{i}||\mathcal{P}^{l}_{j}|}}\;.
    \label{eq:similarity}
\end{equation}
If the intersection of the two sets is empty, then $\theta^l_{i,j}=0$; if the two sets are identical, then $\theta^l_{i,j}=1$.

The resulting overlaps are shown in \autoref{fig:overlap}c, separately for each layer.
First the overlap overall increases up to layer 2 and then clearly drops towards the output layer, indicating that the stages of the paths become increasingly separate from each other.
Remarkably, this structure emerges spontaneously through the learning, with the loss function based on the spike times of the output neurons.
This indicates that the progressively separated paths would be optimal for routing activity towards a specific destination as fast as possible.
Furthermore, this structure would also ensure non-overlapping representations of various input classes towards deeper layers. 

\subsection*{Activity propagates along paths}
\begin{figure*}
    \centering
    \includegraphics[width=\textwidth]{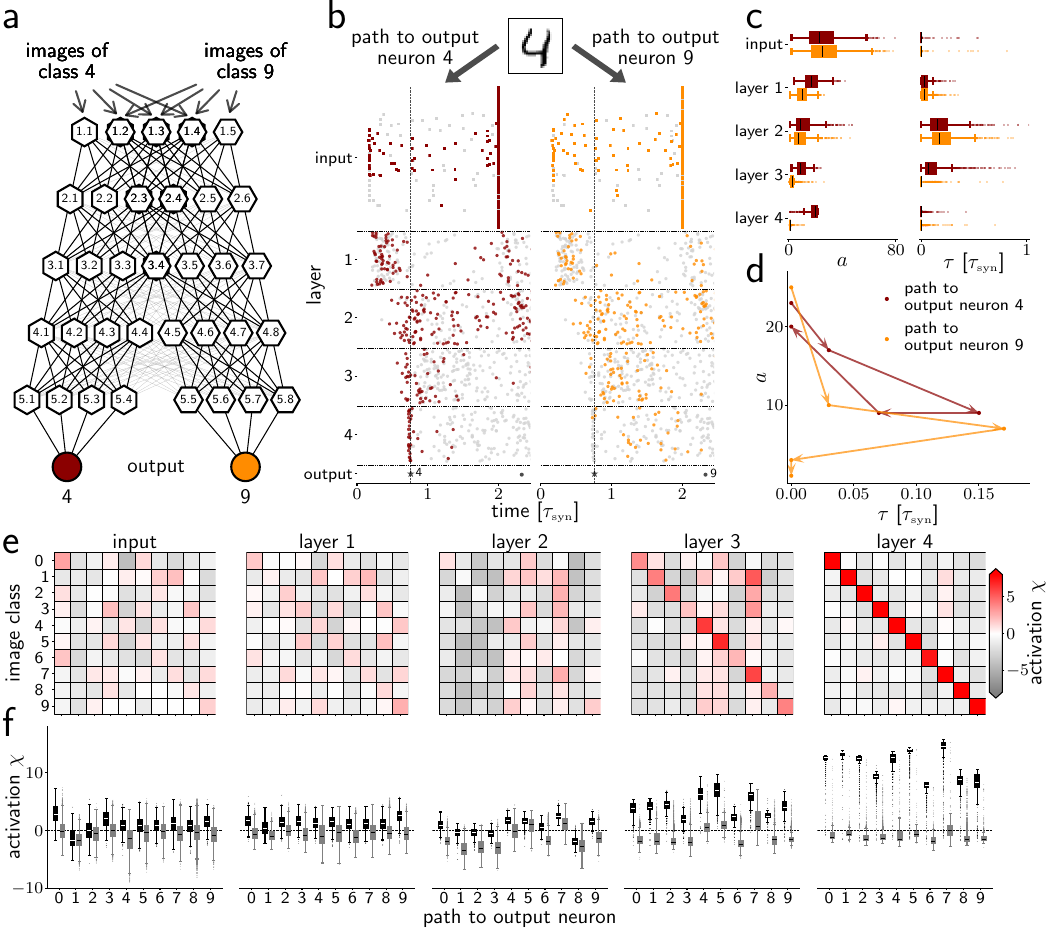}
    \caption{\textbf{Activity propagation along the identified paths.}
             (\textbf{a}) Sketch of the separation of the path to output neuron 4 (\tikzHexTurned) and path to output neuron 9 (\tikzHex) in the deeper layers  of the network.
             (\textbf{b}), Two raster plots (as in \autoref{fig:network_activity}b) of the activity in all layers in response to the same image (class 4) with the neurons labeled according to $\mathcal{P}_{4}$ (dark red) and $\mathcal{N}_{4}$ (gray) (left) and $\mathcal{P}_{9}$ (dark orange) and $\mathcal{N}_{9}$ (gray) (right).
             Inhibitory neurons not shown.
             (\textbf{c}) Box plots of the distributions across all images from class 4 of the number of active neurons $a$ (left) and rise time $\tau$ (right) for the neurons in $\mathcal{P}_{4}$ (red) and $\mathcal{P}_{9}$ (orange) analogous to \autoref{fig:network_activity}e but this time only for a subset of neurons.
             (\textbf{d}) Corresponding state space representation of the medians of the distributions for the two paths in \textbf{d} in sequence of the layers, analogous to \autoref{fig:network_activity}d.
             (\textbf{e}) Evaluation of the activation of the paths.
             Each matrix entry is the mean activation of the paths $\chi$ (\autoref{eq:overlap}) for all images of a given class .
             The colorbar indicates the mean activation, red indicates a strong activation of $\mathcal{P}^{l}_{o}$, gray indicates predominant activation of $\mathcal{N}^{l}_{o}$.
             (\textbf{f}) Box plots for the distribution of the activation for all images, split between the activation when the image class and output neuron are the same (black) or different (gray) for each layer.
             An activation of zero corresponds to chance level (dashed black line).
             }
    \label{fig:dynamic}
\end{figure*}

After the separate analysis of activity and structure, we combine the two.
We ask to what extent connectivity corresponds to dynamics, i.e., how the identified paths relate to the activity patterns discussed before.
As depicted in \autoref{fig:dynamic}a, images of class 4 should activate the neurons in $\mathcal{P}_4$ and their activity should propagate along this path, and the same should hold for images of class 9 and $\mathcal{P}_9$.
This would naturally explain the observed specificity of the leading neurons in their response to images of various classes.

In \autoref{fig:dynamic}b, we show the spikes of excitatory neurons in the same network activity as shown in \autoref{fig:network_activity}b, but here the spikes are labeled according to their membership in two different paths -- on the left: path $\mathcal{P}_4$ to output neuron 4 (red: $\mathcal{P}_{4}$, gray: $\mathcal{N}_{4}$), and on the right: path $\mathcal{P}_9$ to output neuron 9 (orange: $\mathcal{P}_{9}$, gray: $\mathcal{N}_{9}$).
Note that both panels show the same spiking activity, merely labeled differently with regard to different paths.
Specifically, the left panel highlights the spikes through the path to the correct output neuron, and the right panel to an incorrect output neuron.
We observe more and earlier spikes for neurons belonging to $\mathcal{P}_{4}$ compared to those in $\mathcal{P}_{9}$, most evidently in the deeper layers of the network, before the first spike occurs in the output layer.
Furthermore, the spikes of the neurons in $\mathcal{P}_4$ are precisely synchronized in the deeper layers.  
In contrast, the neurons in $\mathcal{P}_9$ emit only a small amount of asynchronous spikes deeper into the network before the first spike in the output layer.

This relates to our earlier observation, namely, the shaping and propagation of pulse packets through the network (\autoref{fig:network_activity}c-f).
Hence, we employ here again the same quantification of the activity using the rise time $\tau$ and the number of leading neurons $a$, but in this case separately for the neurons in each individual path.
The distributions of $a$ and $\tau$ across all images of class 4 for the two paths considered above (\autoref{fig:dynamic}c, $\mathcal{P}_4$ in red and $\mathcal{P}_9$ in orange) show that for both paths the rise time first increases and then decreases, similarly to the previous observation from all excitatory neurons together (\autoref{fig:network_activity}e, red arrows).
However, the two paths behave differently regarding the number of leading neurons.
For $\mathcal{P}_4$, the number of active neurons first decreases and then increases, again as observed in \autoref{fig:network_activity}e (red).
In contrast, for $\mathcal{P}_9$, the number of active neurons decreases constantly, which in part explains the decrease of $\tau$ in the deeper layers in this path, i.e., there are hardly any leading neurons and hence the rise time is bound to be extremely short.
When an image of class 9 is given as the input, we observe the exact inverse (\SIref{fig:SI_images_9}), with activity along path $\mathcal{P}_9$ being shaped into a compact and stable pulse, and the activity along $\mathcal{P}_4$ gradually fading out.
Thus, the present example clearly shows that the activity through the ``correct'' pathway, i.e., the one corresponding to the correct output neuron, survives and gets strengthened, and the activity in the incorrect path dies out (\autoref{fig:dynamic}d).

For the quantification of this observation we define the activation $\chi_{o, x}^l$ of the stage $\mathcal{P}^l_{o}$ in response to image $x$ in layer $l$ as 
\begin{equation}
    \chi_{o,x}^l = \frac{|\mathcal{V}_x^{l} \cap \mathcal{P}^{l}_{o}| - |\mathcal{V}_x^{l} \cap \mathcal{N}^{l}_{o}| - \mu^l_{o,x}}{\sigma^l_{o,x}}\;,
    \label{eq:overlap}
\end{equation}
where we again consider the set of leading neurons $\mathcal{V}_x^{l}$ as defined earlier.
This measure evaluates if the neurons within the path $\mathcal{P}^l_{o}$ or those outside the path $\mathcal{N}^l_{o}$ are more activated.
$\mu^l_{o,x}$ and $\sigma^l_{o,x}$ are for normalizing $\chi_{o, x}^l$ (for details see \methref{sec:meth_overlap}) such that $\chi_{o,x}$ equals zero when the neurons are randomly active independently of their assignments to the path.
The larger $\chi_{o,x}^l$ is, the more neurons within the path are activated compared to the neurons outside the path.
We obtain, for each image $x$, in total ten $\chi_{o,x}^l$ per layer, one for each output neuron.

In \autoref{fig:dynamic}e, we show the mean activation for each of the ten paths across all images of each individual class as a matrix: rows for image classes and columns for output neurons; one matrix per layer.  
The early layers do not show a concentration of activity on the correct path: the diagonal elements of the matrices up until layer 2 are not distinguishable from the off-diagonal elements.
In layer 3 and 4, each path is activated in response almost only to the images of its corresponding class, indicating that in these layers the correct path is selectively activated, with the incorrect paths activated only at a chance level ($\chi\approx0$).
Notably, some paths are in general more active to all images (i.e., the columns for these paths are ``more'' red) than the others, but deeper in the network the activation is highest for images of the correct class.
This is quantitatively confirmed by the distributions of activation (\autoref{fig:dynamic}f) across all images, shown separately for the correct paths (black) and the incorrect paths (gray).
The activation gets increasingly higher for the correct paths deeper in the network, whereas the activation of the incorrect paths stays around zero throughout the layers.

A high activation indicates that the neurons preferentially propagating activity through the path are earlier active than neurons that would propagate activity not through the path.
The activation of the individual path needs to be high enough so that the activity further propagates along that path.
The deciding factor for the selection of the correct path is not the activation of the path compared to the other paths, but whether the activation of the given path is sufficient to further propagate activity. 

\section*{Discussion}

We analyze the spiking activity and connectivity structure of a deep SNN with distinct excitatory and inhibitory populations, trained to classify visual input.
In response to different images, we observe pulse packets i.e., synchronous volleys of spikes, propagating through the network that first broaden and then sharpen again.
While these pulse packets propagate downstream through the network, the neurons active within the packet become increasingly indicative of the image class.
This in particular holds true for the excitatory neurons, while inhibitory neurons generally respond in a less specific manner.
Turning to the network connectivity, starting from individual output neurons we identify different paths each of which corresponds to one image class.
Comparing these paths reveals an increasing separation with network depth on a structural level.
Connecting the analysis of the spiking activity with the identified paths, we demonstrate that upon presentation of an image the evoked activity propagates along the path to the class of the presented image.

Feed-forward networks supporting the propagation of a pulse packet have been discussed extensively in the context of \textit{synfire chains} (SFCs) \parencite{Abeles82, Abeles91, Diesmann99_529, Tetzlaff02_673, Kumar08_1, Trengove13_185}.
SFCs were suggested as a model for reliable and fast propagation of activity in neural networks.
Each of the paths in the network studied here that are activated upon presentation of the various images can be identified as a SFC.
Thus, images are classified by triggering activity along the SFC corresponding to the correct class.
In this sense, the studied network can be thought of as computing with SFCs.
 
Given the large number of neurons in the network, it seems plausible that many more SFCs can be embedded than required to classify MNIST.
In practice, we expect that the number of paths depends on the one hand on the statistics of the input data, i.e., the inter- and intra- class variability, and on the other hand the capacity limit to embed paths in the network \parencite{Bienenstock95_179}.
The evoked activity by different images of the same class needs to converge to the same path while activity for images from different classes needs to be distinguishable. 

Each image class is represented by a distinct subset of neurons that consistently spike early upon the presentation of an image of a given class.
With this, the latency code representing the image is transformed into a binary code of the leading neurons representing the image class.
The neurons in the deeper layer are specialized, i.e., clearly representative of a particular class.
This is similar to the well-known ``grandmother neurons'' or concept cells in areas higher in the visual hierarchy \parencite{Kobatake94_856, Quiroga05_1102, Rust10_12978, Quiroga12_587}.
The representation of the image class becomes clearer with network depth, denoising and semantizing the input through the propagation of signals along the paths \parencite{Kadmon16, Zajzon23_e77009}.
Thus our network reproduces a prominent characteristic of neurons in the visual hierarchy.

Remarkably, the network was not trained with this mechanism in mind: the loss function is based on the spike times of the output neurons and trained the network with regular error backpropagation (for details see \textcite{Goeltz21_823}).
Images were provided in form of spike times with a latency code, an efficient and easy-to-implement code for rapid processing \parencite{Thorpe01_715}.
The described mechanism automatically emerged through the training.
We view this as a direct consequence of the interplay between the spike latency coding in the input, the loss function that enforces competition between the output neurons for who spikes first, and the learning algorithm which ultimately moves spike times to produce the desired outcome.

The network analyzed in this study forms a structure that enables the fast propagation through multiple layers of a network.
Visual processing in the brain shares a similar property: it is well known that in the human brain the visual processing from image presentation to recognition is very fast ($\sim\SI{150}{\milli\second}$) \parencite{Thorpe96_520, Hung05_863}.
Additionally, simple object recognition often relies on the first feed-forward sweep of activity \parencite{Lamme00_571, Roelfsema23_1003}, and information is transmitted by the first spikes in response to a stimulus \parencite{Johansson04_170}.
The processing in our network relies also only on the feed-forward sweep of activity.
This is sufficient for classifying MNIST.
For more complex object recognition \parencite{Kar19_974} or other cortical processes, like attention \parencite{Lamme00_571, Super01_304} or learning \parencite{Hinton95_1158} recurrent connections are suggested to be required.
The influence of recurrent connections on the here studied mechanism needs to be studied in future work.

By including excitatory and inhibitory neurons, we recover another property observed in cortical networks: excitatory neurons are more sharply tuned to a specific stimulus, while the inhibitory neurons are less specific \parencite{Sohya07_2145, Niell08_7520, Lundqvist10_e803}.
Inhibitory neurons are employed during the propagation of the activity, but they do not carry the main information about the image class.
Rather, they regulate the network by providing unspecific inhibitory input to the excitatory neurons in the next layer, akin to the \textit{blanket of inhibition}, i.e. the dense and unspecific innervation of excitatory neurons by inhibitory neurons, found in cortical circuits \citep{Fino11_1188, Karnani14_96}.
Similarly, inhibition has been found to restrict the spatial spread and temporal persistence of neural activity in visual cortex \parencite{Haider13_97}.
Additionally, inhibition could facilitate the synchronization of the pulse packets, as had been reported in a previous study \parencite{Shinozaki10_011913}. 
In our network the inhibitory neurons develop a similar facilitating role though the training.

We note that the size of the employed network consists of a much larger number of neurons and layers than necessary to classify MNIST.
In the original implementation, \textcite{Goeltz21_823} showed that the task can already be solved  by a network with only one hidden layer.
Since in this work we aimed at investigating the relation between signal propagation and computation, we chose a network that contained more layers.
We expect the result to be transferable to more complex visual tasks, since MNIST does not contain any structure that inherently enforces the observations we report here. 

Recently the theoretical analysis of the dynamics of learning capabilities in artificial neuronal networks
has gained attention \parencite{Schoenholz17_01232, Fischer23_arxiv_07715, Vanmeegen25_3345}.
The approaches in these studies allow for a statistical assessment across different networks.
In contrast, here we focused the analysis on one concrete realization of an SNN.
This complementary approach enables a dissection of the relationship between structure and function on a more fine-grained level, doing justice to the individuality of each trained network.
However, results for other realizations are qualitatively similar (\SIref{fig:SI_other_network}).
Focusing on a specific network acknowledges the fact that natural neural networks are not the averages of a distribution, but a concrete instance that grow and adjust to fulfill a specific function. 
With our \textit{idiographic} \parencite{Windelband98_5} approach we provide insight into SNNs, even if we base the analysis on only few examples.
In this way, our approach is similar to the analysis of neuroscientific experiments, where one also has access to only a few subjects \parencite{Fries22_1114}. 

Future work could address how different spike timing codes, imposed by construction, would shape the learned activity in the network.
Similarly, the impact of different learning rules could be analyzed.
In this way, the universality of the identified shared properties between the visual system and the networks studied here can be investigated.
Additionally, a thorough analysis of SNNs may help to also improve their performance \parencite{Dold25_arxiv}.

Expanding the approach of our analysis to more complex networks and  more complex visual tasks will strengthen the connection between functional neural networks and fundamental concepts in neuroscience.
This includes whether trained SNNs form receptive fields, or if through the training binding emerges \parencite{Singer95_555}.
Moreover, future analyses could also address networks with recurrent connections and ongoing activity.
Thus, we view this work explicitly as a starting point for further studies of how structures inside the brain are capable of learning efficient spike-based codes.
While our comparatively simple networks already allow the formulation of clear and rigorous links between gradient descent on spike times and observations in cortex, further extensions of our model will provide additional insight into the computational role of the various components -- in structure and dynamics 
-- observed in the brain.
    
\section*{Methods}\label{sec:Methods}

\subsection*{Network setup}

The investigated networks are multi-layer, feed-forward, all-to-all connected networks of spiking neurons.
Here, we elaborate on the setup of the experiments (for details see~\cite{Goeltz21_823}):
The neurons are leaky integrate-and-fire neurons with exponential synapses and a long refractory time constant to ensure single spikes per neuron.
Following an input sample, the spiking activity of the neurons is given by a differentiable function, and its derivatives are used to optimize the parameters in the network with gradient descent (Equations~2, 4, 5 in~\cite{Goeltz21_823}) in a mini-batch training setup.
The precise parameters of training as well as the training code are given alongside the trained network, see \nameref{sec:code}.

In a change from the referenced manuscript, here we respect Dale's law and separate the neurons into an excitatory and an inhibitory population in the hidden layers.
We ensure the desired effect by clipping the outgoing weights to positive and negative values, respectively.

\subsection*{Rise time}\label{sec:meth_rise}
The rise time $\tau_x$ is measured on the basis of the population spike time histogram in each layer individually for the activity in response to image $x$.
For the calculations in this paper, we calculate the spike time histogram with a sliding bin size of $0.05\:\tau_{\text{syn}}$ with non-exclusive binning.
The rise time is defined as the center of the first bin that corresponds to a maximum after the first spike in the layer.

\subsection*{Set of leading neurons}\label{sec:meth_active}
We define a set of leading neurons $\mathcal{V}_x^l$ for image $x$ in layer $l$ on the basis of the rise time $\tau_x$.

\begin{align}
    \mathcal{V}_x^l =  \{i|t_{i,x} <= \min_{i}(t_{i,x}) + \tau_x\}\;,
\end{align}
with the spike time $t_{i,x}$ of each neuron $i$.

\subsection*{Similarity of sets of leading neurons}\label{sec:meth_corr}
To measure similarity between the sets, we define a measure on the basis of the Pearson product-moment correlation coefficient.
For that, we need to define a mean $\mu$, a variance and covariance in the context of our sets.
To this end we interpret the neurons of in a layer as a binary vector with $N^l$ elements $v_{x,i}^l$ for which $v_{x,i}^l=1 \: \forall \: i\in\mathcal{V}_x^l$ and $v_{x,i}^l=0 \: \forall \: i \notin\mathcal{V}_x^l$.
In this framework we use the mean over the entries of this vector.:
\begin{align}
    \mu_x^l = \frac{1}{N^l}\sum_{i=1}^{N^l}v_{x,i}^l = \frac{|\mathcal{V}_x^l|}{N^l}\;.
\end{align}
Accordingly, for the variance we have:
\begin{align}
    \text{Var}_x^l & = \frac{1}{N^l} \sum_{i=1}^{N^l}(v_{x,i}^l-\mu^l_x)^2 \\& =  \frac{1}{N^l} \sum_{i=1}^{N^l}\left((v_{x,i}^l)^2 - 2v_{x,i}^l\mu^l_x + (\mu^l_x)^2\right) \\ & = \frac{1}{N^l} \left(\sum_{i=1}^{N^l}(v_{x,i}^l)^2\right) - (\mu^l_x)^2 \\& = \frac{|\mathcal{V}_x^l|}{N^l} - \left(\frac{|\mathcal{V}_x^l|}{N^l}\right)^2\;,
\end{align}
and the covariance: 
\begin{align}
    \text{Cov}_{x,y}^l & = \frac{1}{N^l} \sum_{i=1}^{N^l}(v_{x,i}^l-\mu^l_x)(v_{y,i}^l-\mu^l_y) \\ & = \frac{1}{N^l} \left(\sum_{i=1}^{N^l}v_{x,i}^lv_{y,i}^l\right) - \mu^l_x\mu^l_y \\ & = \frac{|\mathcal{V}^l_{x} \cap \mathcal{V}^l_{y}|}{N^l} - \frac{|\mathcal{V}^l_{x}||\mathcal{V}^l_{y}|}{(N^l)^2}\;.
\end{align}
From this we obtain the similarity between sets as defined in \autoref{eq:correlation}:
\begin{align}
    \rho^l_{x,y} & = \frac{\text{Cov}_{x,y}^l}{\sqrt{\text{Var}_x^l\text{Var}_y^l}} \\ & = \frac{N^l|\mathcal{V}^l_{x} \cap \mathcal{V}^l_{y}| - |\mathcal{V}^l_{x}||\mathcal{V}^l_{y}|}{\sqrt{|\mathcal{V}^l_{x}||\mathcal{V}^l_{y}|(N^l - |\mathcal{V}^l_{x}|)(N^l - |\mathcal{V}^l_{y}|) }}\;.
\end{align}

\subsection*{Information gain}\label{sec:meth_IG}
We first measure the entropy H of the distribution of image classes $X$. we then measure the neuron-conditional entropy $H(X|i)$, i.e., the entropy of the posterior distribution of $X$ given that neuron $i$ was active.
The information gain $\text{IG}_i$ for neuron $i$ is the normalized difference between these two entropies:
\begin{align}
      \text{IG}_i = \frac{H(X) - H(X|i)}{H(X)}.
\end{align}

\subsection*{Assignment of neural subsets and paths}\label{sec:meth_subset}
For identifying the paths, we start with the neurons in the last hidden layer, since they are directly responsible for the classification by the output neurons.
Let's consider output neuron $o$.
To evaluate which neurons in layer 4 preferentially target this neuron, we compare the connection weight $w_{o, i}^4$ of each neuron $i$ to the output neuron $o$ with the average weight of given neuron to all output neurons: $\bar{w}_{i}^4 = \frac{1}{10}\sum_{j=0}^{9} w_{j, i}^4$ with respect to the standard deviation $\sigma^4_{w_i}$ of these weights.
All neurons in layer 4 that fulfill $w_{o, i}^4 > \bar{w}_{i}^4 + \sigma^4_{w_i}$ are assigned to the set $\mathcal{P}^4_{o}$ of neurons in layer $4$ with strong impact on output neuron $o$.
The neurons that do not fulfill this condition are in set $\mathcal{N}^4_{o}$, resulting in:
\begin{align}
    \mathcal{P}^4_{o} = & \{i|w_{o, i}^4 > \bar{w}_{i}^4 + \sigma^4_{w_i}\} \\
    \mathcal{N}^4_{o} = & \{i|w_{o, i}^4 \leq \bar{w}_{i}^4 +  \sigma^4_{w_i}\},
\end{align}
with $\sigma^4_{w_i} = \sqrt{\frac{1}{10}\sum_{j=0}^{9} (w_{j, i}^4 - \bar{w}_{i}^4)^2}$.

Then in the penultimate layer (layer $3$), we calculate for each neuron the average connection weight to neurons in $\mathcal{P}^4_{o}$ and $\mathcal{N}^4_{o}$, respectively:
\begin{align}
    \bar{w}^{3}_{i, p, o} = & \frac{1}{|\mathcal{P}^{l}_{o}|} \sum_{j\in\mathcal{P}_o^4}w^{3}_{j, i} \text{ and} \\ 
    \bar{w}^{3}_{i, n, o} = & \frac{1}{|\mathcal{N}^{l}_{o}|} \sum_{j\in\mathcal{N}_o^4}w^{3}_{j, i}.
\end{align}
On this basis we again assign neurons to two sets:
\begin{align}
    \mathcal{P}^{3}_{o} = & \{i|\bar{w}^{3}_{i, p, o} > \bar{w}^{3}_{i, n, o}\}\\
    \mathcal{N}^{3}_{o} = & \{i|\bar{w}^{3}_{i, p, o} < \bar{w}^{3}_{i, n, o}\}.
\end{align}
This procedure is repeated backwards through the whole network, until all excitatory neurons in the hidden layers and the neurons in the input layer are assigned.

\subsection*{Activation of neural subsets}\label{sec:meth_overlap}
The activation $\chi_{o,x}^l$ as defined in \autoref{eq:overlap} is normalized with respect to random activity of the neurons.
It is used to evaluate whether $|\mathcal{V}_x^{l} \cap \mathcal{P}^{l}_{o}|$ or $|\mathcal{V}_x^{l} \cap \mathcal{N}^{l}_{o}|$ is larger.
For this it takes into account the expected value $\mu_{o, i}$ and the standard deviation $\sigma_{o, i}$, if the leading neurons would be drawn randomly from $\mathcal{P}^{l}_{o}$ or $\mathcal{N}^{l}_{o}$ respectively.
\begin{align}
    \mu_{o, i} = & |\mathcal{V}^{l}_{i}| \left(2\frac{|\mathcal{P}^{l}_{o}|}{N^l} - 1\right) \\
    \sigma_{o, i} =& \sqrt{4\frac{|\mathcal{P}^{l}_{o}|}{N^l} \left(1-\frac{|\mathcal{P}^{l}_{o}|}{N} \right) |\mathcal{V}^{l}_{o, i}|\frac{N-|\mathcal{V}^{l}_{i}|}{N-1}}
\end{align}
The probability of drawing a neuron from $\mathcal{P}^{l}_{o}$ is $\frac{|\mathcal{P}^{l}_{o}|}{N^l}$.
Accordingly, the probability for drawing a neuron from $\mathcal{N}^{l}_{o}$ is $\frac{|\mathcal{N}^{l}_{o}|}{N^l} = 1 - \frac{|\mathcal{P}^{l}_{o}|}{N^l}$.
We draw $|\mathcal{V}^{l}_{i}|$ neurons without replacement from the layer.
This corresponds to a hyper-geometric distribution, thus follows $\mu_{o, i}$ as mean and $\sigma_{o, i}$ as standard deviation if the neurons were drawn randomly.

\subsection*{Code and data availability}\label{sec:code}
Code for the network simulations is available at \url{https://github.com/JulianGoeltz/fastAndDeep}.
Code for the analysis will be made available as of the date of publication. 
Any additional information required to recreate the results reported in this paper is available from the lead contact upon request.

\section*{Author contributions}
Conceptualization, JOF, JI, MAP, SG;
Methodology, JG, LK, MAP;
Software, JOF, ACK, JG, LK;
Formal Analysis JOF, ACK, JI, SG;
Investigation, JG, LK, MAP;
Visualization, JOF, ACK;
Writing – Original Draft, JOF, ACK, JG, LK, JI, MAP, SG;
Writing – Review \& Editing,  JOF, ACK, JG, LK, JI, MAP, SG;
Funding Acquisition, MAP, SG;
Resources, MAP, SG;
Project Administration, JOF, SG;
Supervision, JI, MAP, SG. 

\section*{Acknowledgements}

    We thank Markus Diesmann, G\"unther Palm and Shigeru Shinomoto for intense and fruitful discussions. This research was funded by the European Union’s Horizon 2020 Framework programme for Research and Innovation under Speciﬁc Grant Agreements No. 945539 (HBP SGA3) and No. 101147319 (EBRAINS 2.0 Project), the NRW-network 'iBehave'(NW21-049) and the Helmholtz Joint Lab SMHB. This work was performed as part of the Helmholtz School for Data Science in Life, Earth and Energy (HDS-LEE). We further wish express our gratitude to the Manfred Stärk Foundation for their continuing support of the NeuroTMA Lab.

\printbibliography

\onecolumn

\section*{Supplemental Information}\label{sec:SI}
\FloatBarrier

    \setcounter{figure}{0} 
    \renewcommand{\thefigure}{S\arabic{figure}} 
    \renewcommand{\theHfigure}{S\arabic{figure}} 
    
    \begin{figure*}[htb]
        \centering
        \includegraphics[width=\textwidth]{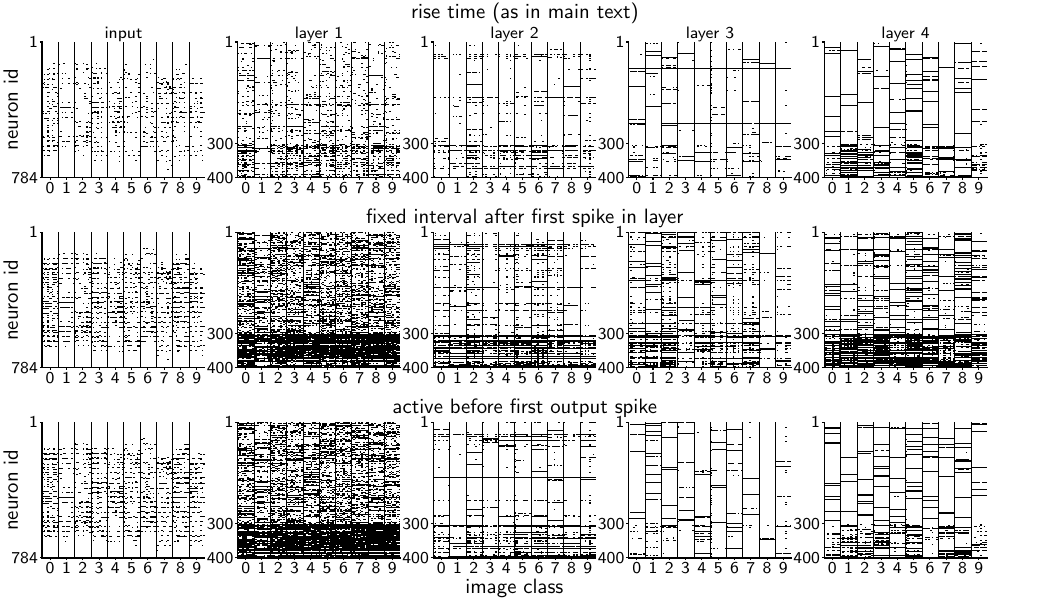}
        \caption{\textbf{Robustness for the definition of the sets of leading neurons.}
        We can define the set of leading neurons in different ways: in the first row we show result on the basis the definition as in the main text, in the middle row the set is defined as all neurons that spike in the first $0.5\:\tau_{\text{syn}}$ after the first spike in the layer and in the last row the set is defined as all neurons that spike before the first output spike.
        }
        \label{fig:SI_other_def}
    \end{figure*}
    
    \begin{figure*}[htb]
        \centering
        \includegraphics[width=\textwidth]{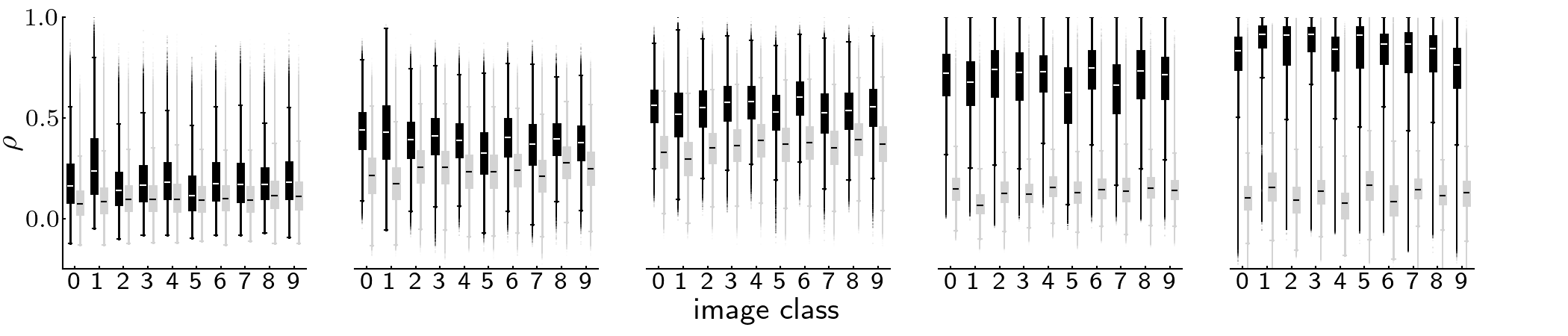}
        \caption{\textbf{Distribution of similarities.}
        Here we show the distribution of similarities across all images, split between image pairs of the same class (black) and different classes (grey).
        }
        \label{fig:SI_corr_dist}
    \end{figure*}
    
    \begin{figure*}[htb]
        \centering
        \includegraphics[width=\textwidth]{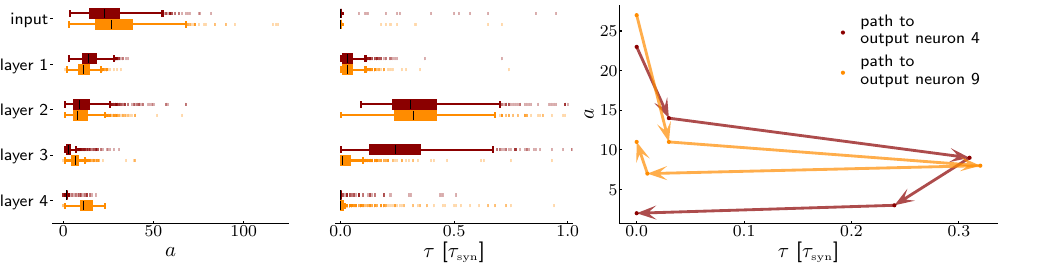}
        \caption{\textbf{Robustness of the result for images from another class.}
        The results in this figure correspond to the result shown in \autoref{fig:dynamic}c and \autoref{fig:dynamic}d, but for all images from class 9.
        }
        \label{fig:SI_images_9}
    \end{figure*}
    
    \begin{figure*}[htb]
        \centering
        \includegraphics[width=\textwidth]{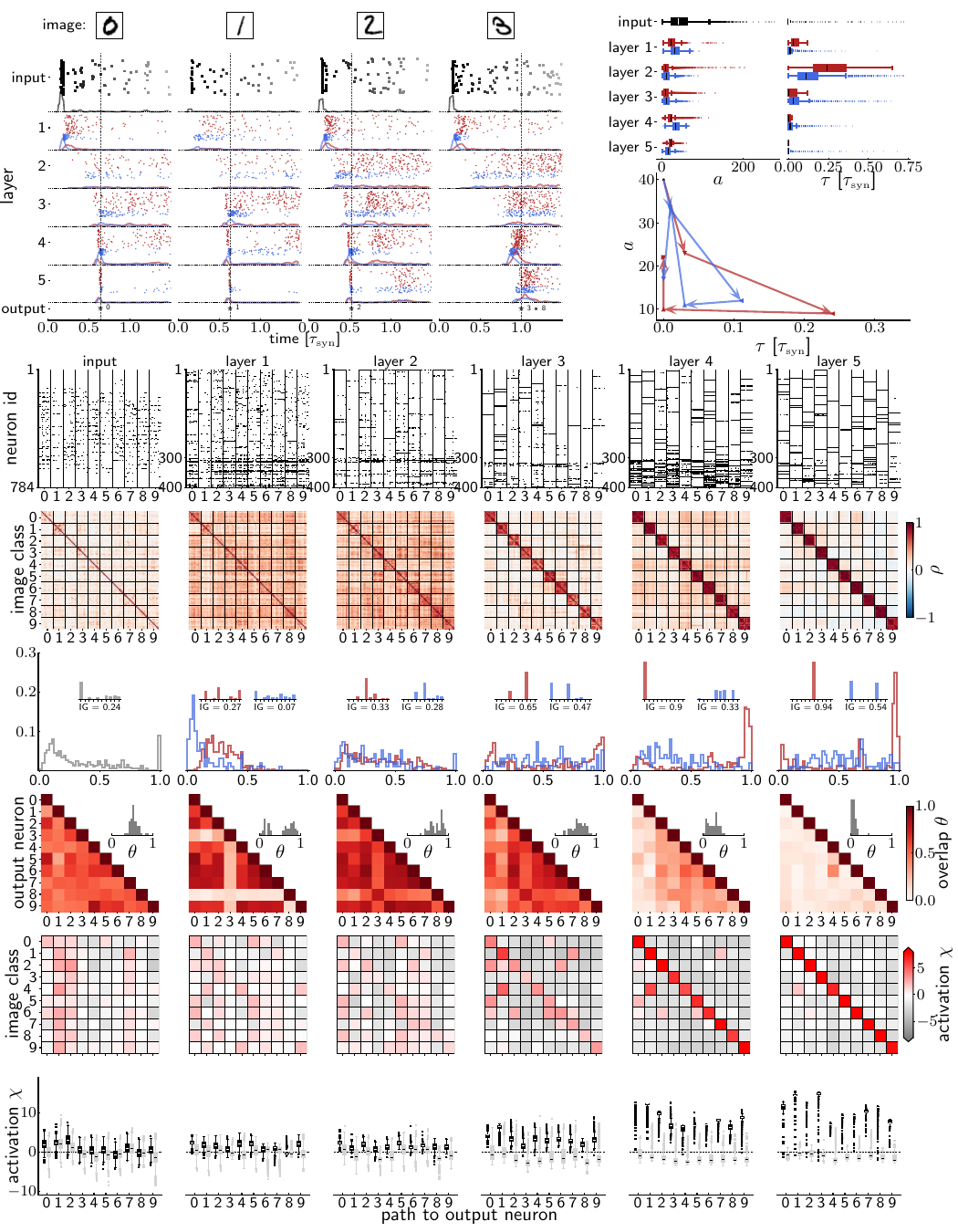}
        \caption{\textbf{Summary figure for all results obtained from network initialized with another seed and 5 hidden layers in total.}
        The results are analogous to the corresponding figures in the main document.
        }
        \label{fig:SI_other_network}
    \end{figure*}

\end{document}